# Limiting the Impact of Light Pollution on Human Health, Environment and Stellar Visibility

## Authors and Affiliations

Fabio Falchi and Pierantonio Cinzano
*Istituto di Scienza e Tecnologia dell'Inquinamento Luminoso, Via Roma 13, I-36106 Thiene, Italy*

Christopher D. Elvidge
*NOAA National Geophysical Data Center, Boulder, Colorado USA*

David M. Keith
*Marshall Design Inc., Boulder, Colorado US*

Abraham Haim
*The Israeli Center for Interdisciplinary studies in Chronobiology, University of Haifa, Haifa 31905, Israel*

Corresponding author: F. Falchi; e-mail: falchi@lightpollution.it,

## Abstract

*Light pollution is one of the most rapidly increasing types of environmental degradation. Its levels have been growing exponentially over the natural nocturnal lighting levels provided by starlight and moonlight. To limit this pollution several effective practices have been defined: the use of shielding on lighting fixture to prevent direct upward light, particularly at low angles above the horizon; no over lighting, i.e. avoid using higher lighting levels than strictly needed for the task, constraining illumination to the area where it is needed and the time it will be used. Nevertheless, even after the best control of the light distribution is reached and when the proper quantity of light is used, some upward light emission remains, due to reflections from the lit surfaces and atmospheric scatter. The environmental impact of this "residual light pollution", cannot be neglected and should be limited too. Here we propose a new way to limit the effects of this residual light pollution on wildlife, human health and stellar visibility. We performed analysis of the spectra of common types of lamps for external use, including the new LEDs. We evaluated their emissions relative to the spectral response functions of human eye photoreceptors, in the photopic, scotopic and the 'meltopic' melatonin suppressing bands. We found that the amount of pollution is strongly dependent on the spectral characteristics of the lamps, with the more environmentally friendly lamps being low pressure sodium, followed by high pressure sodium. Most polluting are the lamps with a strong blue emission, like Metal Halide and white LEDs. Migration from the now widely*

*used sodium lamps to white lamps (MH and LEDs) would produce an increase of pollution in the scotopic and melatonin suppression bands of more than five times the present levels, supposing the same photopic installed flux. This increase will exacerbate known and possible unknown effects of light pollution on human health, environment and on visual perception of the Universe by humans. We present quantitative criteria to evaluate the lamps based on their spectral emissions and we suggest regulatory limits for future lighting.*

## Introduction

Light pollution is the alteration of natural light levels in the night environment produced by introduction of artificial light. Due to the continuous growth of nighttime artificial lighting, this problem is increasingly debated and many localities have developed regulations to constrain the wasteful loss of light into the sky and environment.

The expanding use of light at night is because humans are diurnal animals that are trying to extend activities into the usually dark hours. This increasing use is driven by what seems common sense, and by the lighting industry with justifications that at first may seem correct. With few exceptions, everything we build is lit at night. This includes streets, roads, bridges, airports, commercial and industrial buildings, parking lots, sport centers and homes. Outdoor lighting continues to expand as more infrastructure is built. Lighting levels are often set high with one or more secondary objectives in mind. For instance, building exteriors are often lit for a merely aesthetic effect. Shopping centers are typically heavily lit to attract shoppers and create a lively environment designed to stimulate spending. Lighting levels in public areas are often set high as a deterrent against crime, even though studies have not proven this to have any effect on crime rates [1-3]. Indeed the cores of our urban centers are bathed in light and the resulting light pollution can extend more than a hundred kilometers out from the city's edge.

There is reliable evidence that this artificial extension of the day produces serious adverse consequences to, human health and the environment.

The impact of light pollution on the night sky has been described in depth by Cinzano, Falchi and Elvidge [4]. In the First Atlas of Artificial Night Sky Brightness they showed that more than 60% of world population lives under light polluted skies (99% of the population of USA and Europe) and almost one-fifth of world terrain is under light polluted skies.

In regards to human, to date there are no doubts that exposure to light at night (LAN) decreases pineal melatonin (MLT) production and secretion and are not only a source for phase shift in daily rhythms. Apart of timing and exposure duration, the two light variables responsible for the suppression of MLT production are: 1) light intensity and 2) wave length. Therefore, it seems that the combination of both variables should be considered for the threshold of LAN. Light intensity levels found to suppress MLT production are decreasing as research progresses. During the eighties of last century, it was shown that bright light at an order of thousands of lux was requested for abolishing the secretion [5]. The discovery of a novel photoreceptor, the Non Image Forming Photoreceptors (NIFPs), and the photopygment melanopsin gave an opportunity for a better understanding of light perception by humans and showed the effects of spectrum in the human high response to LAN exposure [6-11]. The results of a study [12], in which the impact of wave length on humans was assessed by measuring melatonin, alertness, thermoregulation and heart rate draw the attention to the significant role of wave length. It was shown that exposure of two hours to monochromatic light at 460nm in the late evening significantly suppressed melatonin secretion while under the same intensity, exposure timing and duration but at wavelength of 550nm such effects were not observed. Already Wright et al. [13] showed that even illuminance as low as 1.5 lux affect circadian rhythms. However, recently it as shown that bedroom illumination, typical of most homes in the evening, is sufficient to reduce and delay MLT production [14]. From the results of these studies it can be noted that MLT suppression by LAN is wavelength depended and intensities can be much lower than those used several decades ago.

Alteration of the circadian clock may cause performance, alertness, sleep and metabolic disorders. Exposure to light at night suppresses the production of the pineal hormone melatonin, and since melatonin is an oncostatic or anti-carcinognenic agent, lower levels in blood may encourage the growth of some type of cancers [15-20]. MLT seems to have an influence on coronary heart disease [21]. LAN acts directly on physiology, or indirectly by causing sleep disorders and deprivation, that may have negative effects on several disorders such as diabetes, obesity and others [22-23]. For a brief review of physiological, epidemiological and ecological consequences of LAN see Navara and Nelson [24].

Therefore, the increase in light intensity on the one hand and the wide use of "environmentally friendly bulbs" with a short wave length emission on the other, are probably having sever negative impact on health through the suppression of MLT production.

In the natural environment, animals and plants are exposed to light at night levels that vary from about $5 \times 10^{-5}$ lux of the overcast sky, to $1 \times 10^{-4}$ lux by the starry sky on a moonless night, to $2 \times 10^{-2}$ lux at the quarter moon, to 0.1-0.3 lux during the week around full moon. The artificial light of a typical shopping mall, 10-20 lux, is up to 200 thousand times brighter than the illuminance experienced in the natural environment around new moon. No wonder that it has become apparent that light at night has strong environmental effects in behavioral, population and community ecology (in foraging, mating, orientation, migration, communication, competition, and predation) and effects on ecosystems. For a review of ecological consequences of light pollution see [24-28]. This strong evidence of the adverse effects of artificial light at night on animals and on human health should be balanced against the supposed positive effects on safety and security. Fortunately it is possible (and also simple in theory, if those involved in lighting collaborate) to limit the light pollution effects and, at the same time, allow for the lighting that is usually perceived as a need by people. Practical ways to limit the effects of light pollution on the night sky and the night environment are well known and verified [29]:

a) **Full Cutoff Shielding:** Do not allow luminaires to send any light directly at and above the horizontal, with particular care to cut the light emitted at low elevations (in the range gamma=90-135 degrees above the downward vertical, i.e. 0-45 degrees from the horizon plane). In practice, light in this range travels long distances through the atmosphere and enhances the additive property of light pollution [30-31], an effect that compounds the problem, especially in densely populated areas. An additional limitation on the light leaving the fixture downward (in the range gamma=80-90 degrees from the downward vertical, i.e. 0 to 10 degrees below the horizon plane) should also be enforced. This is because the nearly-specular reflection of asphalt at grazing incidence considerably increases the amount of light at low angles above the horizontal (although this reflected light is much more subject to screening by surrounding vegetation and buildings). This limitation will also improve the comfort and visual performance of road users by lowering the direct glare from fixtures.

b) **Limiting the Area of Lighting**: Carefully avoid wasting downward light flux outside the area to be lit. Such waste is not only a main cause of increase of installed flux per unit surface (and in turn a main cause of increase in energy expense), but some of this light is also reflected upward from these surfaces. Even if Lambertian diffusion from horizontal surfaces is less effective in sending light at low elevations than direct emission by luminaires, nevertheless when the direct emission is eliminated, the diffuse reflection remains as an appreciable source of pollution.

c) **Eliminate Overlighting:** Avoid luminances or illuminances greater than the minimum required for the task, and dim lights when the application allows it.

d) **Shut Off Lights When Not in Use[1]:** It makes sense to turn the lights off when you leave the room, or for the lights to turn off automatically, but in outdoor lighting these options are rarely available (in Italy, for example, almost all the parking lots of shopping malls are lit all night long, and likewise for the industrial/artisan/commercial areas, whether or not there are workers at night).

e) **Limit Growth in Installed Lighting**: Limits to the increase of the new installed flux should be implemented. A 1% yearly increase could be allowed at first for each administrative area, followed by a halt in the increase of total installed flux, and then by a decrease. This does not mean that no new installation will be allowed, but simply that if you want to install new lights you have to decrease the flux in the existing overlighted areas.

To these basic prescriptions, some others could substantially improve lighting quality (e.g. a requirement that the lighting installation be designed by a professional lighting designer, although this might not be feasible in poorer countries nor advisable for smaller installations, provided they respect the code) or to take account of specific kinds of installations (e.g. signs or historical buildings). Most of these prescriptions are already implemented in some of the most advanced anti-light- pollution laws such as Lombardia (Italy) Regional Law n.17 of March 27, 2000 with its subsequent additions and modifications. Twelve other similar regional laws followed in Italy, and most Italian territory and population are now protected by these laws. Slovenjia adopted a similar law in year 2007. Falchi [32] found that despite an almost doubling in the outdoor installed flux, in two studied sites in Lombardia, the artificial sky brightness did not increase over the last twelve years. This is probably due to the adoption of laws against light pollution in the surroundings of the sites. A full enforcement of the prescriptions could probably make a substantial improvement in the quality of the night sky and environment. In fact, the same research shows that in six studied sites, on average, 75% of the artificial sky brightness is produced by light escaping directly from fixtures and only 25% from the reflections off lighted surfaces. This implies that, all the rest being equal, a complete substitution of the installed fixtures with fully shielded ones could lower the artificial sky brightness to 1/4 of present levels. In two of the studied sites, more than 90% of the artificial sky brightness derived by direct light. These sites would presumably have a 90% decrease in light pollution as a result of retrofitting fixtures to fully shielded in the surrounding territory that produce light pollution, i.e. a circle of at least 100 km radius.

---

[1] Even a great reduction (1/10$^{th}$ of the full values) of lighting levels could be advised, but safety norms don't allow for this.

Nevertheless, even when the best control of the light distribution is reached and when the proper quantity of light is used, some upward light emission remains, due to reflection from the properly lighted surface. This is an unavoidable by-product of the lighting operation, even when measures a), b) and c) have been achieved: lighting is installed just to produce reflections of light. However, after the light has performed its useful function, it is then dispersed into the environment. Due to its near-Lambertian behavior, this reflection is frequently less effective at low elevations than at large elevations, so the effect on the night sky tends to be confined largely to the vicinity of the source. In any case, the environmental impact of this residual light pollution cannot be neglected.

Limitation of this residual pollution requires limits not only on "how" nighttime lighting is arranged according to prescriptions a) and b), but also "how much" nighttime lighting is made. Typically it has been proposed to limit the growth rates of installed flux in each city, or to limit the average density of installed light flux (e.g. installed flux per hectare or acre). However, following the example of the radio portion of the electromagnetic spectrum, there is an additional way to limit this residual pollution: by preferential use of light sources with spectral characteristics that have the least impact on star visibility and human and wildlife health, while maintaining a given degree of visibility in areas that need artificial lighting. This would allow reduction of the negative astronomical and biological effects without impairing essential night lighting.

This solution has been applied for decades whenever Low Pressure Sodium (LPS) and High Pressure Sodium (HPS) lamps have been requested in place of Mercury Vapor (MV) or incandescent lamps. The arrival of new LED light sources for night-time outdoor lighting and widespread use of broad spectrum Metal Halide (MH) lamps even where they aren't the best option enhances the need to define a more quantitative prescription, applicable to any kind of lamp and capable of giving precise indications to the lighting industry on the way to proceed in light source development or improvement (e.g. how to filter or tailor the spectrum of the emitted light).

The prescription should:
(i) be as effective as possible in protecting the night environment from the adverse effects of light pollution;
(ii) take account of existing night time lighting habits in order to minimize the impact on human activities:
(iii) allow easy identification of non-compliant light sources; and
(iv) allow easy measurement in the field, when needed.

In this paper we discuss the problem, we recognize two different quantitative parameters, we devise a prescription and we investigate how it could be enforced.

## Methods

The possibility of limiting the residual light pollution, avoiding the need to limit night time outdoor lighting itself, is based on the different response with wavelength of the two main classes of eye receptors and the action spectrum of circadian rhythm disruption for rodents, monkeys and humans [33]. In a schematic way and for the purposes of this paper, we can distinguish the photopic response of cones and the scotopic response of rods. The eye response is fully photopic, i.e. cones fully determine it, for luminances over 3 cd/m$^2$ whereas the eye response is fully scotopic, i.e. rods fully determine it, under about 0.01 cd/m$^2$. In the range between these two limits, called mesopic, the eye response goes from scotopic to photopic, depending on the relative contributions of the two classes of receptors, which in turn depend on the luminances in view. (Figure 1).

Standard rules, e.g those on road safety lighting, usually require road luminance to be in the range from 0.3 to 2 cd/m$^2$ but, even where laws against light pollution prohibit exceeding values suggested by standard rules, in practice new installations rarely have an average maintained luminance under 0.75 cd/m$^2$, the prescribed luminance of the ME4b class of the European Norm EN 13201. Eye response at these luminances is predominantly photopic[2] (see discussion below). In fact when we look at artificially-lit outdoor areas and recognize colors, which is a property of cones, it indicates that our cone vision is functioning. In some cases colors could be distorted by lamp spectra but in any case they are recognized. Otherwise, we could use monochromatic lamps like Low Pressure Sodium lamps everywhere and there would be no reasons to use white light. Moreover, the 0.3 to 2 cd/m$^2$ prescribed range is for the luminance of the road surface, usually dark asphalt, while the night scene in a city is also full of other lights and surfaces that usually have far higher luminances: the direct lights from fixtures, light colored objects, vehicle lights, billboardsand shop windows. So our eyes are not fully dark adapted in a typical city night scene (see Figure 2).

In observation of the starry sky, where the natural luminance of the sky is about 200 μcd/m$^2$, the response is scotopic, except when looking at a few bright stars. This difference gives us a way of separating the primary polluting effects of the light from its lighting capabilities. Unfortunately the scotopic and photopic response curves overlap in part, as shown in Figure 3. This prevents us from fully separating these two effects. This means that we cannot use the spectra of lamps to limit light pollution in place of full–cutoff fixtures and the other prescriptions. Even monochromatic lamps emitting at the maximum of photopic response, e.g. LPS lamps, contribute consistently to the scotopic response pass band. So the a), b) and c) prescriptions listed in the introduction are still required in practice. However we can use this differential response to diminish pollution, subject to all the usual precautions to limit the amount of light pollution.

At luminances under 0.5 cd/m$^2$ the contribution of monochromatic rods to eye response could be relatively larger, but we are aware that there are very few new installations with average maintained luminance so low in Italy, even if standard rules allow 0.3 cd/m$^2$ for local roads. The Lighting Research Center recommends using the usual photopic values at luminance higher than 0.6 cd/m$^2$ [34]. Lewis [35] claimed that under a photopic luminance of 0.1 cd/m$^2$ some MH lamps emitting strongly in the scotopic passband can produce a reaction time only slightly worse than that obtained with some HPS lamps at 1 cd/m$^2$. In a study on mesopic visibility Orreveteläinen [36] investigated reaction times in seeing different colour targets in peripheral vision. He found no differences at 1 cd/m$^2$, very small differences at 0.1 cd/m$^2$ and evident differences only at 0.01 cd/m$^2$., where the blue and cyan targets were detected earlier than warmer colour targets. Lewis [35] showed that for a luminance of 1 cd/m$^2$ MH lamps emitting consistently in the scotopic band still produce a slightly greater contrast than HPS. However, better visibility in the peripheral field at the edge of streetlight obtained with bluer light should be evaluated along with the strong decrease in eye lens transparency of blue wavelengths with age. Brainard et al. [37] found that at 450 nm the transmittance of the lens of 60-69 year old is half that of 20-29 year old adults. At 425 nm it is one third. At 555 nm it is only a few per cent less while it is equal at 600 nm and above. Studies on vision should use a variety of subjects of different ages, to take into account the increasing population of elder drivers. A migration toward bluer lamps, such as MH and LEDs, will exacerbate the difference in vision performance between young drivers and old drivers, penalizing the latter even as they become a greater fraction of the driving population.

---

[2] Even the Lighting Research Center recommends using the usual photopic lumen at luminance greater than 0.6 cd/m$^2$ instead of their proposed Unified Luminance that take into account for the blue content of the lamps in the mesopic range [24].

One additional observation should be made concerning high-blue-content lamps. Road surface materials, either asphalt and concrete, reflect less short-wavelength radiation compared to long-wavelength radiation, as seen in figures 4 and 5. This implies that lamps that emit more long-wavelength radiation, such as LPS or HPS, will have more light reflected by these roads. On the other hand, lamps that emit more in shorter-wavelengths are less effective in producing luminance from these road surfaces. We computed that at equal photopic output white LED lighting produces 6% to 11% less luminance from roads than HPS , depending on the type of surface. The spectra of these lamps are shown in figure 6. The spectral reflectivity of roads reduces the blue contribution of the lamps by one half or more, lowering the effects on the environment but also lowering the supposed visual benefits. Moreover, if fixtures are not suitably shielded, this lowering in the reflection of blue light is negligible, due to the dominant contribution of direct light to sky luminance outside of cities.

Replacement of HPS lamps with MH lamps and white LEDs - with an accompanying reduction of luminance to 0.1 cd/m$^2$ - does not seem immediately applicable because (a) existing rules do not allow such small luminances and typically require a very time-consuming process to be changed, (b) existing studies do not seem sufficiently complete and convincing to justify these practices.
As new studies on the negative effects of artificial light at night will be produced, a lowering of the external lighting levels would probably be advisable even in the case of a demonstrated decrease of visibility on roads. Accumulated evidence of the demonstrated negative effects of light at night may well outweigh the positive ones. Moreover, most of the positive effects used to justify the huge expenses to build, maintain and power external lighting are based on anecdotal indications or poor statistical analysis [1-3].

Even for the road safety effect there is a lack of studies using randomised controlled trials. A public registration of protocols and trials is suggested, lowering the problem of publication bias [38] by ensuring that 'against lighting' results remain as visible as 'for lighting' ones.

**The wavelengths that cause the worst light pollution**

For nearly a hundred years the specification and characterization of light has been based on the wavelength-dependent sensitivity of the two recognized types of photoreceptors (rods and cones) in the human eye. Rods are solely responsible for scotopic or night vision which is black and white. The cones are solely responsible for color vision. There is a wavelengthsensitivity to the vision provided by the rods or cones. The photopic band is a spectral representation of the sensitivity of the just the cones, and is centered in the green portion of the spectrum. When under dim lighting conditions there is insufficient light for activation of any cones, the rods are still able to provide black and white vision. This is scotopic vision, which has peak sensitivity in the blue-green (Figure 3). Early in this decade [6-10] a third photoreceptor in the human eye was recognized – a circadian photoreceptor with wavelength sensitivity centered in the blue [6,7]. As noted earlier, exposure to lighting with a high blue component disrupts the normal melatonin rhythms, commonly leading to insomnia, stress and increased risk for a wide range of medical maladies and even cancer. Preventing the blue component from reaching the eye by means of filters blocking wavelengths under 530 nm, preserves nocturnal melatonin production in humans [11]. This implies that the blue component of light has the severest consequences for the environment and human health.

A second reason that blue light contributes more to light pollution than green or red light is that blue light is more readily scattered in the atmosphere – as you can see from the blue sky of daylight hours. This "blue sky" effect arises from Rayleigh scattering which is inversely proportional to the fourth power of wavelength, meaning that shorter wavelength radiation - blue light - will scatter in the atmosphere more than longer wavelength radiation - green and red light [39]. The Sun at sunset and sunrise appears orange because the blue component of its light has been redirected by the atmosphere. But this style of scattering also applies to light emitted by cities and towns at night. Green and red light emitted upward are scattered less than blue light, so a higher portion of the long wavelength light tends to continue on towards space. More of the blue light is scattered in the atmosphere, contributing to the sky brightness that we call light pollution.

**Scotopic to Photopic Ratio**
The first way to minimize the impact of residual light pollution is to use lamps that, for a given amount of photopic light flux, produce a minimal amount of scotopic light flux. In fact, lighting installation design and standard rules are based on photopic luminance. At parity of photopic performance, we can sacrifice the small and usually unknown scotopic contribution, and in exchange we gain the chance of lowering the light pollution effects substantially.
The photopic luminous flux is defined as [40]:

$$\Phi_V = K_m \int_0^\infty \Phi_{e,\lambda} V(\lambda) d\lambda \qquad (1)$$

where $V(\lambda)$ is the photopic response [41], $\Phi_{e,\lambda}$ is the spectral radiant flux of the source and $K_m$=683 lm/W is the photometric efficacy, i.e. the standard lumen per watt conversion factor, for photopic response [42].
The scotopic luminous flux is defined as [40]:

$$\Phi_{V'} = K'_m \int_0^\infty \Phi_{e,\lambda} V'(\lambda) d\lambda \qquad (2)$$

where $V'(\lambda)$ is the scotopic response [43] and $K'_m$=1699 lm/W is the photometric efficacy, i.e. the standard lumen per watt conversion factor, for scotopic response [44].
Then the scotopic-to-photopic ratio $R_{sp}$, commonly used in lamp performance comparisons, is:

$$R_{sp} = \frac{\Phi_{V'}}{\Phi_V} = \frac{K'_m \int_0^\infty \Phi_{e,\lambda} V'(\lambda) d\lambda}{K_m \int_0^\infty \Phi_{e,\lambda} V(\lambda) d\lambda} \qquad (3)$$

It gives the scotopic light flux of a lamp for unit photopic light flux. The ratio of radiant fluxes in the scotopic and photopic spectral ranges will generally vary and are of little value for the present purpose.
We used standard CIE responses, neglecting more recent and accurate photopic responses known as Judd [45] modified $V(\lambda)$, Judd-Vos modified $V_M(\lambda)$ [46] and Stockman and Sharpe [47] $V^*_2(\lambda)$, because for now standardization of the ratio has priority over accuracy.
Setting an upper limit on the scotopic/photopic ratio could help to control or prevent the strong growth of artificial night sky scotopic luminance that would be produced by a migration from the current population of HPS lamps to MH or LED lamps promoted by the lighting industry because of their white output.

**Protected spectral band for visual astronomy**
Due to the above mentioned overlap of the photopic and scotopic luminosity curves, minimizing the scotopic to photopic ratio might not provide enough protection for the night sky in the short wavelength part of the visible spectrum. Hence a more specific wavelength-based restriction on the emissions from lighting is appropriate.
Let's consider a hypothetical lamp with a given ratio of scotopic to photopic light flux and a given radiant flux in each of the two bands. Now let's assume that we are able to move the spectral flux emitted in the wavelength range 440-540 nm to the blue side of the scotopic band, below 440 nm. Let's finally assume that we are able to tune this flux and the remaining photopic flux in order to maintain the same scotopic and photopic fluxes as before so that the eye will perceive the same quantity of light in scotopic and photopic pass bands. The color of the lamp will change slightly, due to the shift. However now the range 440-540 nm is much darker. The artificial night background produced by the considered lamp will be negligible when observing the night sky with a filter that blocks any wavelength outside this range.

Given that stellar visibility in unpolluted conditions is limited by eye sensitivity, it is necessary that the "protected" wavelength range be centered on the maximum of the scotopic response curve and as large as possible (at least 100 nm) in order that the impact of the filter on limiting stellar magnitude be kept as small as possible. Otherwise the reduction in the eye's scotopic sensitivity with the filter will annul any advantage of filtering out the artificial part of skyglow.

We choose the scotopic protected interval, hereafter called P-band, in the range 440-540 nm in place of the range 450-550 nm in order to leave the mercury emission line at 546 nm unaffected. This range bounds 79 per cent of the area under the scotopic curve.

In practice, we cannot "move" light of a lamp toward redder or bluer wavelengths but we can make a good start by using lamps that have a relatively weak output in the P-range. We also need to devise a way to filter out the light in the 440-540 nm range of every lamp, e.g. using absorbing pigments on the lamp glass or on the fixture's cover glass. This could lead to design of a lamp (i) with whiter light than HPS thanks to emission lines in the blue that tend to balance the emissions over 540 nm; (ii) leaving the peak of the scotopic response unpolluted; and (iii) maintaining the same scotopic-to-photopic ratio as HPS lamps of today.

Lumen is not defined in bands different than scotopic and photopic, so we need to define a parameter in terms of energy flux.

The radiant flux in the photopic band $\Phi_{e,V}$ is:

$$\Phi_{e,V} = \int_0^\infty \Phi_{e,\lambda} V(\lambda) d\lambda \qquad (4)$$

where $V(\lambda)$ is the photopic response of CIE and $\Phi_{e,\lambda}$ is the spectral radiant flux of the source.

The radiant flux $\Phi_{e,P}$ in the protected band $\lambda_0$-$\lambda_1$ is:

$$\Phi_{e,P} = \int_{\lambda_0}^{\lambda_1} \Phi_{e,\lambda} d\lambda \qquad (5)$$

The P-band radiant flux to photopic luminous flux ratio $R_P$, hereafter called P-ratio, is:

$$R_P = \frac{\Phi_{e,P}}{\Phi_{e,V}} = \frac{\int_{\lambda_0}^{\lambda_1} \Phi_{e,\lambda} d\lambda}{\int_0^\infty \Phi_{e,\lambda} V(\lambda) d\lambda} \qquad (6)$$

It gives the energy emitted in the "protected" band by a lamp emitting unitary luminous flux in the photopic response pass band and, in practice, measures the lamp impact on the protected band.

An effective upper limit on the P-ratio is a practical way of protecting stellar visibility from harm caused by artificial radiant flux in the spectral range 440-540 nm. Following this approach could at least make the wavelength range near the maximum of scotopic sensitivity minimally polluted (figure 4).

## Measurements

Emission spectra were acquired using an ASD, Inc. FieldSpec 3 spectroradiometer equipped with an 8 degree field of view foreoptic. The instrument had been radiometrically calibrated and spectra were acquired in radiance (W/m$^2$/µm/sr) mode over the 350 to 2500 nm range. Each lamp was warmed up prior to measurement and the spectra were acquired from one lamp at a time in a dark room. The measured light sources included the following classes 1) liquid fuel lamps, 2) pressurized fuel lamps, 3) incandescent, 4) quartz halogen, 5) metal halide, 6) high pressure sodium, 7) low pressure sodium, and 8) light emitting diodes (LED).

Given the limited number of lamps and manufacturers in the market, the P-ratios should be provided by manufacturers for each lamp, calculated from the spectral power distribution, or measured. It would be futile to leave this to lighting installers to do. A quick check of installed lamps in the field by competent technicians will determine if the type of lamp installed satisfies the P-ratio limit discussed below.

The scotopic to photopic ratio can be obtained by dividing the illuminances measured using luxmeters with interchangeable filters available on the market. A scotopic luxmeter can also be purchased directly or obtained by replacing the photopic filter in a suitable luxmeter with a scotopic filter and calibrating it. An energy measurement of the P-ratio can be obtained with an irradiance-meter provided with an interference filter for the range 440-540 nm, and calibrated in irradiance with the filter in place. Such filters are commercially available, and some irradiance meters are already provided with photopic and scotopic filters. The calibration can be made using one of the many spectral calibration standards available on the market. There is no need for lighting installers to acquire such equipment provided that manufacturers' data are reliable, but it could be a good idea for environmental control organizations to acquire the equipment.

Table 1 shows actual ratios for some cases of interest in external lighting. Scotopic to photopic ratios have been measured at LPLAB by Cinzano [48], or calculated. Average scotopic to photopic ratios for HPS and MH lamps are taken from Knox & Keith [49]. P-ratios have been computed with synthetic photometry from spectra taken by Cinzano with WASBAM [50] and from spectra measured by Elvidge and Keith.

| **Lamp** | $R_{sp}$ | $R_P$ |
| --- | --- | --- |
| LPS (*) | 0.20 | 0.0027 |
| HPS 70 W (*) | 0.55 | 0.13 |
| Average HPS(**) | 0.66 | - |
| HPL 80 W (Hg vapour) (*) | 1.18 | 0.27 |
| CIE Illuminant A (***) | 1.41 | 0.51 |
| QTH 3100 K (*) | 1.56 | 0.58 |
| Average MH (**) | 1.60 | 0.46 |
| Flat spectrum (***) | 1.86 | 0.93 |
| LED 'natural white' (***) | 3.5 | 0.87 |

Table 1. Ratios for some lamps and lamp classes.
(*) $R$sp measured
(**) $R$sp from Knox & Keith [45]
(***) $R$sp calculated

What upper limits should we set for the previous ratios? In principle, lamps with minimum ratios should be adopted. By far, the least polluting lamps in the P-band are the Low Pressure Sodium, with a $R$p ratio about 2% of the second best lamps, High Pressure Sodium. Unfortunately, LPS lamps have the disadvantages of long length and poor colour rendition, along with diminishing availability. In many applications colour rendition is unimportant or unnecessary, but LPS lamps have been abandoned by the lamp manufacturers in favor of other popular lamp types. An LPS lamp should be first choice, with others used only if strictly necessary. However, following the compromise position of individuals and organizations working against light pollution, we suggest that the upper limits be set equal to the actual maximum ratios of most common HPS lamps. Lamps with still larger ratios should be used only in those cases when strong reasons for "whiter" light are demonstrated. We are in a phase similar to when catalytic converters were introduced in the car market. They didn't stop pollutants totally (as in an ideal world), but started to do so in a way compatible with the technology of the time. Similarly, setting a limit compatible with HPS will start controlling the blue content, while not upsetting the current habits of the market. Due to the overwhelming importance of our health over the "necessity" to use white or blue-rich light, even applications other than road lighting should follow our prescription. These top limits could be enforced by law as obligatory, because voluntary quality goals may not suffice.

Table 1 also shows that a migration from the current population of HPS lamps to MH or, worse, blue-rich white LED lamps could produce a growth of artificial night sky brightness by 2.5 to 5 times, as perceived by the dark adapted human eye. Such large increases, combined with the usual growth of installed flux, may produce a tenfold increase of the scotopic sky brightness in the next ten years or so. Figure 5 shows the spectral power distributions of a white LED and a HPS lamp with equal photopic lumen output. The far higher emission of blue by the LED is evident.

The same order of magnitude increase is expected on the melatonin suppression action spectrum. This action spectrum is shown in Figure 6. The peak sensitivity is from 440 nm to 500 nm. A tenfold decrease in sensitivity over the 460-470 nm peak is shown at the 410 nm and 540 nm wavelength. As seen in Table 2, given the same photopic lumen output, MH and LED lamps emit 3 to 7 times more energy in the 440 nm to 500 nm band compared to HPS. Considering the full range of the action spectrum, the results are similar. Given the uncertainties in the spectrum itself, we can summarize that MH is about three times more polluting in this band than HPS. Natural white LED has more than double the content of MH. A migration from HPS lamps to MH lamps and white

LEDs could produce far worse effects on human health than today's lighting does. This will impair and negate worldwide efforts towards better and less polluting lighting practices. . LEDs have anyway a great potential, they could be tuned and produced with very different spectra, so it is advisable that industry research be pushed toward the production of less polluting warm LEDs, with no blue emissions.

| Lamp type | Energy relative to HPS, 440 to 500 nm band | Melatonin suppression effect (relative to HPS) |
|---|---|---|
| HPS | 1 | 1 |
| LPS | 0.02 | 0.3 |
| Metal Halide | 2.7 | 3.4 |
| Natural White LED | 7.0 | 5.4 |
| Incandescent 65 W | 2.5 | 2.5 |

Table 2: 440 nm to 500 nm energy ratios (second column) and melatonin suppression efficiency (third column) for some common lamps.

## Proposed limits

Residual light pollution is that produced by reflected light, after direct upward emission has been accurately minimized, overlighting has been avoided, and the flux wasted illuminating outside surfaces has been minimized. It would remain to be dealt with after laws or regulations have required zero direct emission above the horizontal by lighting fixtures, limited the luminance or illuminance to the minimum required by security rules, minimized as much as possible the fraction of light wasted downward outside the surface to be lighted, and banned the use of mercury vapour[3] in every type of lamps.

The International Agency for Research on Cancer has recently added to the list of group 2A (probably carcinogenic to humans) shiftwork that involves circadian disruption [51]. As seen, circadian disruption is also induced by light exposure at night and light at night is becoming a public health issue [52, 53].

Light pollution has to be recognized as a hazard to our environment and our health and not, as commonly believed, as just a problem for astronomers. This view is supported by the recent resolution of the American Medical Association [54] where it is said that light pollution is a public health hazard.

**We recommend a total ban of the outdoor emission of light at wavelengths shorter than 540 nm to reduce the adverse health effects of decreased melatonin production and circadian rhythm disruption in humans and animals.** The relatively low emissions of HPS lamps in this spectral range could be set as the limit[4] on what is acceptable in terms of the balance between photopic and meltopic emission ratios. So, this rule should be use as standard:

---

[3] These lamps must be prohibited anyway due to their mercury content and low efficacy

[4] This limit is a compromise due to the available types of lamps on the market. It could be lowered in future, but it is anyway sufficient to stop the growth of the blue light content in the environment due to the LEDs and MH lamps.

*The wavelength range of the visible light spectrum under 540 nanometres, corresponding to high sensitivity of the melatonin suppression action spectrum, should be established as a protected range. Lamps that emit an energy flux in the protected range larger than that emitted by the standard HPS[5] lamp on a basis of equal photopic output should not be installed outdoors[6].*

The following prescription aims to limit residual light pollution in the scotopic band and should be used only in the limited number of cases where there is the absolute necessity to have accurate colour perception and the previous rule cannot be followed:

*The wavelength range of the visible light spectrum between 440 and 540 nanometres, corresponding to the maximum sensitivity of the scotopic vision of the human eye, should be established as a protected range. Lamps should not be installed outdoors if (a)[7] their emission in this wavelength range exceeds 15 per cent of the energy flux emitted in the photopic response pass band, measured in watts, and (b)[8] their emission in the scotopic response pass band exceeds two-thirds of that emitted in the photopic response pass band, measured in lumens.*

In the authors' opinion, lamp producers should follow these rules as a minimum precaution in order to minimize the impact of their products on human health and on the environment, even in the absence of laws or regulations.

Following the actual market trend towards more, brighter and whiter light may expose lamp producers and the lighting industry to extensive litigation for illness caused by toxic products, as has already happened with the tobacco and asbestos industries.

A regulation, to be studied, for lamps for interior use during night could be introduced too.

## Conclusions

In this paper we analysed the different energy and luminous fluxes in the melatonin suppression action spectrum and in the scotopic band for several types of lamps. We found that huge differences exist in the blue emissions of the lamps, for the same photopic luminous flux. Due to the fact that night vision and our health are impaired more by blue light, we proposed two limits to be followed in the adoption of lamps for external use. The first should be used everywhere, as a standard, in order to reduce emissions within the melatonin suppression band at night, as much as possible,. The second rule should be used only in a very limited number of situations where better colour rendition is indispensable for the task.

Therefore, an effective law to control light pollution should implement this set of rules:
- do not allow luminaires to send any light directly at and above the horizontal;
- do not waste downward light flux outside the area to be lit;

---

[5] HPS energy flux varies with the power of the lamp. So, for each lumen output of the lamp to be evaluated, it is to consider the immediate lower power HPS lamp. For example, in evaluating a 14000 lm lamp, it must be compared to a 100 W HPS lamp that typically produces 9500 lm instead of a 150 W HPS lamp that emits about 16000 lm.

[6] Regulamentation of indoor lighting lies outside the purposes of this paper

[7] i.e. Rp ratio lower than 0.15

[8] i.e. Rsp ratio lower than 0.66

- avoid over lighting;
- shut off lights when the area is not in use;
- aim for zero growth of the total installed flux;
- strongly limit the short wavelength 'blue' light.

Application of all these prescriptions would allow for proper lighting of our cities and, at the same time, protect ourselves and the environment from the more adverse effects of light pollution.

## Acknowledgments

We acknowledge Dr. Barry A.J. Clark and Dr. Jan Hollan for very interesting discussions and suggestions and consequent improvement of this work. We acknowledge also Dr. Steven W. Lockley and Dr. Paul Marchant for help given in their respective fields.

Figure Legends

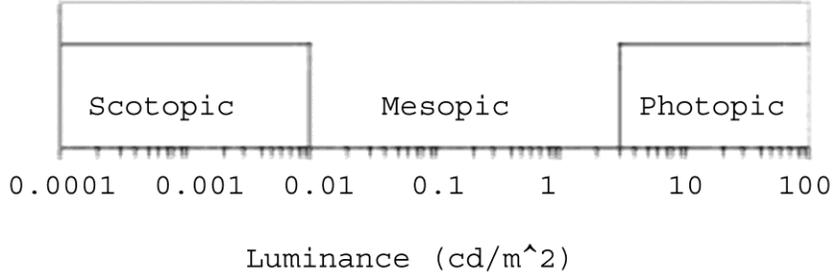

Figure 1. Approximate luminance ranges of scotopic (rods), mesopic (rod/cone transition region) and photopic (cones) response of the human eye.

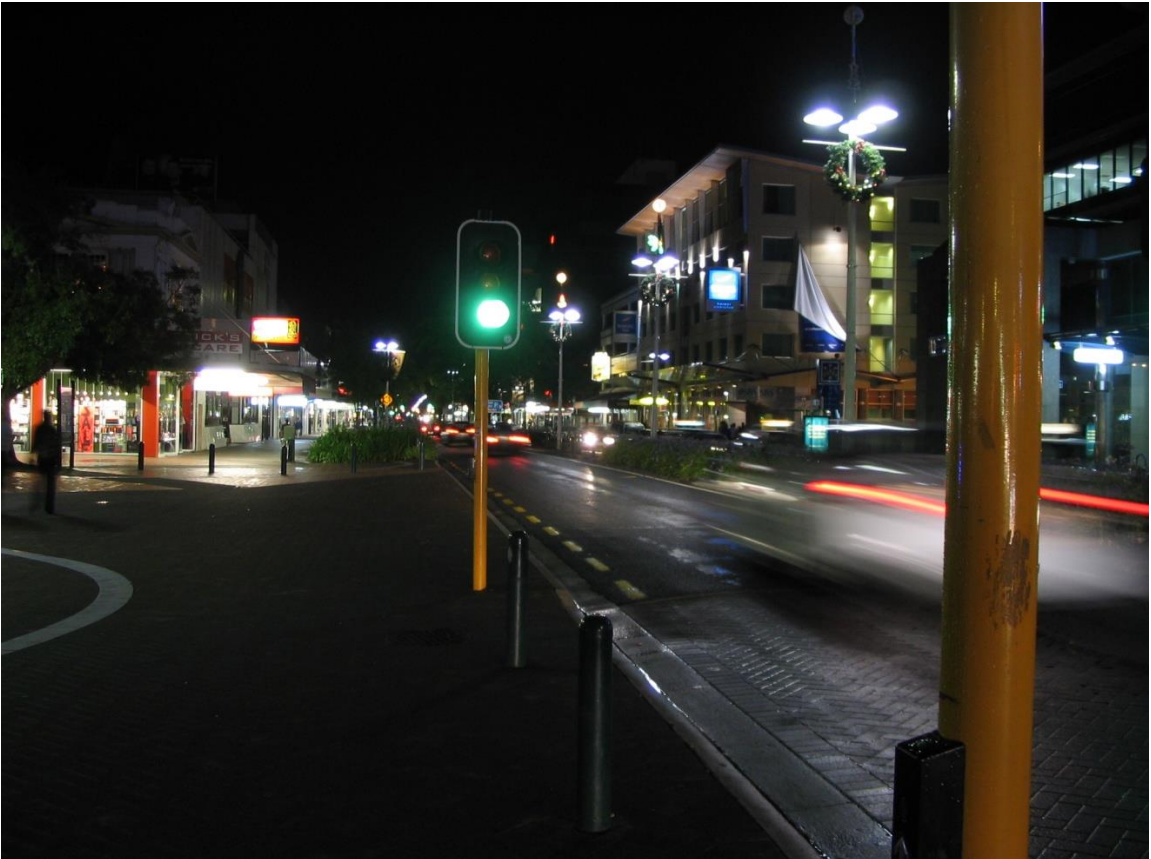

Figure 2. A typical town night scene. The lowest luminance is on the ground and on the street where it should be about 1 cd/m$^2$. Its luminance is even lower than the night sky in big cities. Most of the rest of the scene has a far higher luminance, completely in the photopic range of our eyes. (Photo by Bruce Kingsbury)

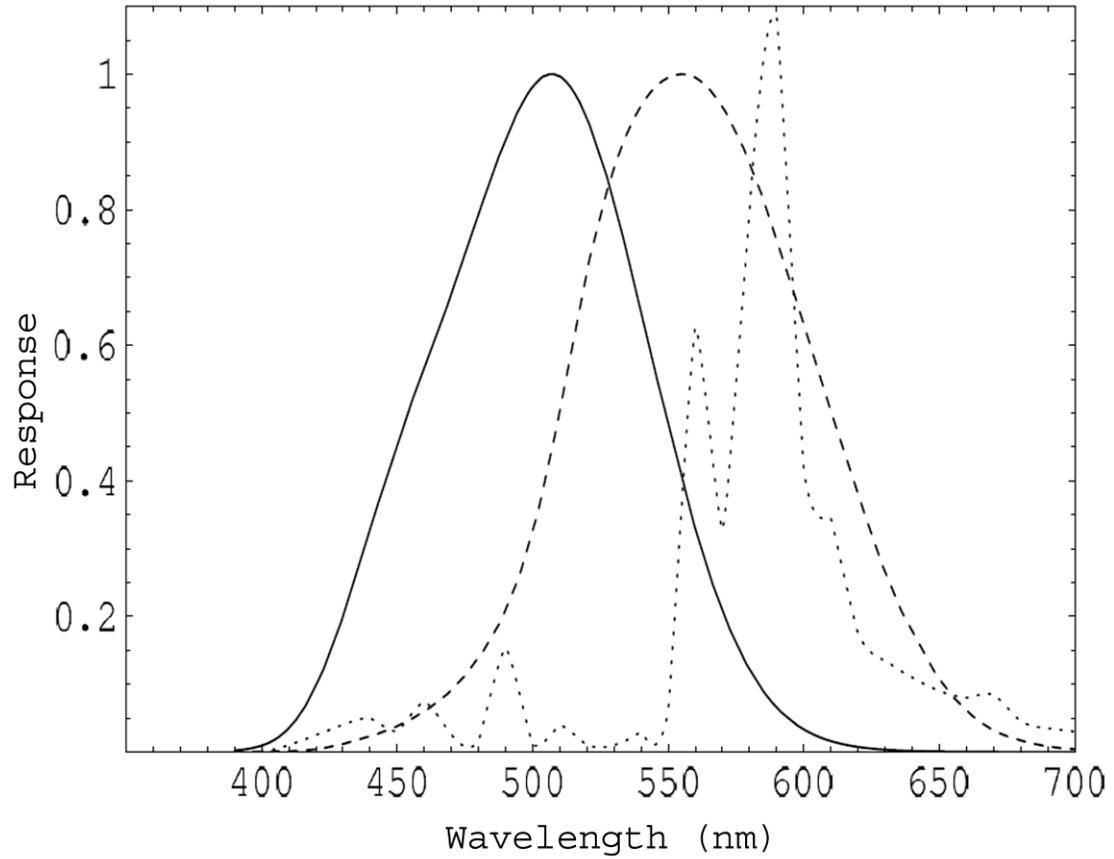

Figure 3. Photopic (dashed line) and scotopic (solid line) normalised responses for comparison with the spectral power distribution of a HPS lamp (dotted line).

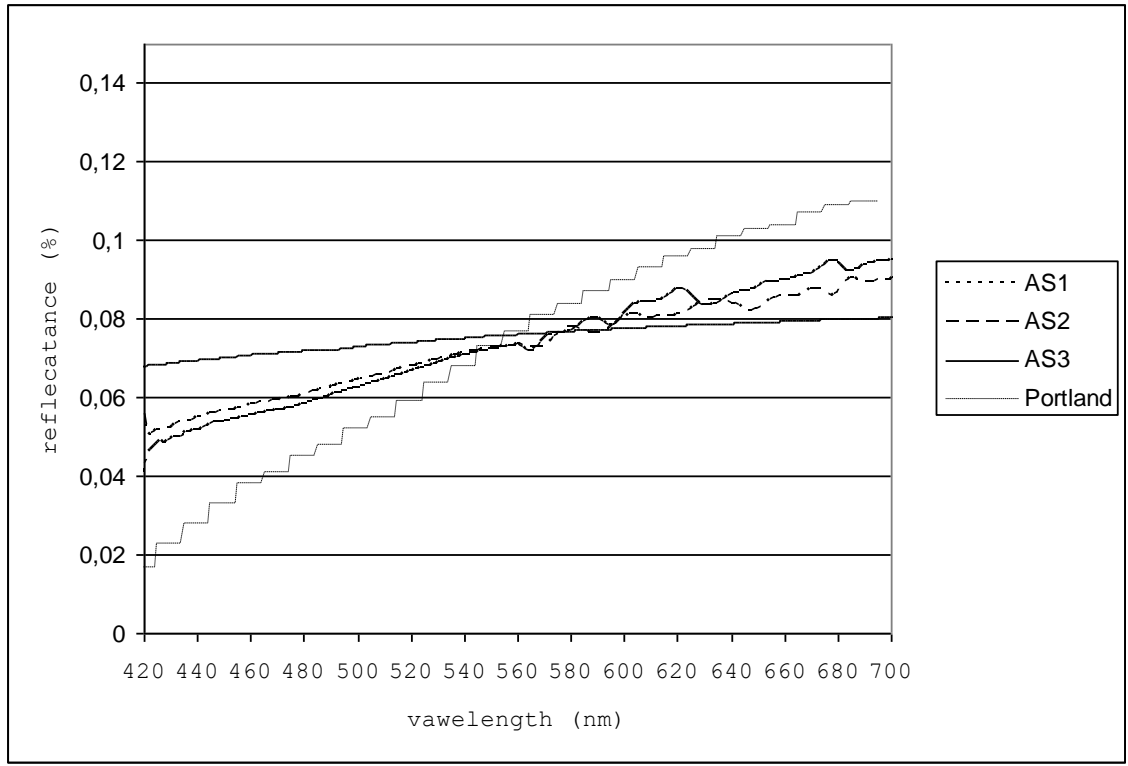

Figure 4. Spectral reflectance of four asphalt surfaces. Data from NASA/Jet Propulsion Laboratory ASTER Library [55] and Portland Cement Association [56]

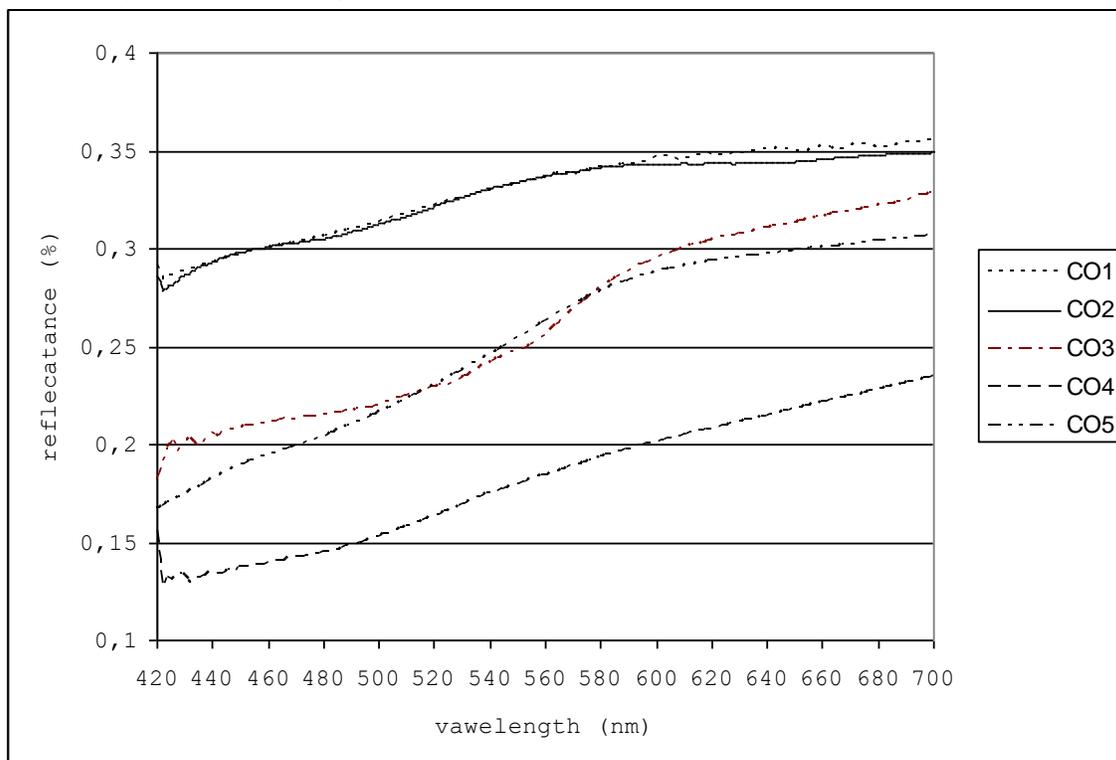

Figure 5. Spectral reflectance of five concrete surfaces. Data from NASA/Jet Propulsion Laboratory ASTER library [55]

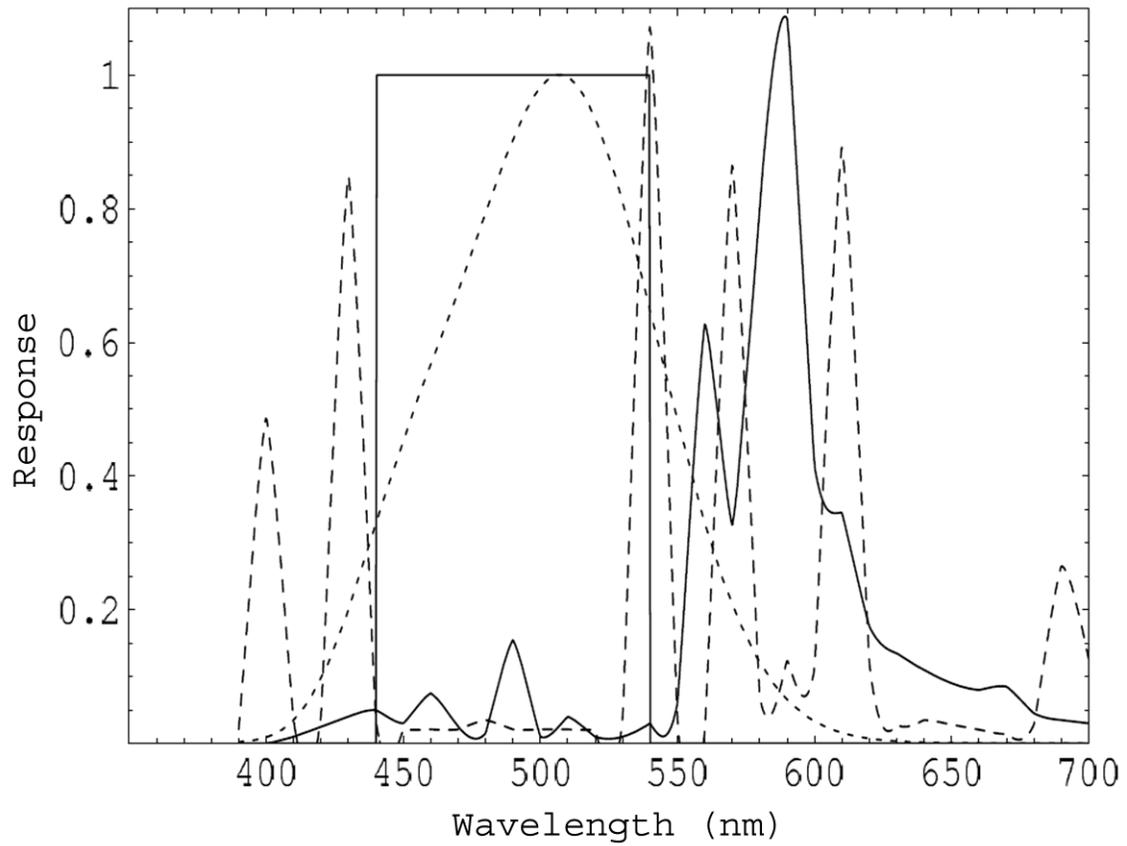

Figure 4. The protected 440-540 nm range (solid line) compared with the scotopic response (short dashed line) and the spectral power distributions of a HPS lamp (solid line) and a Mercury Vapour lamp (dashed line).

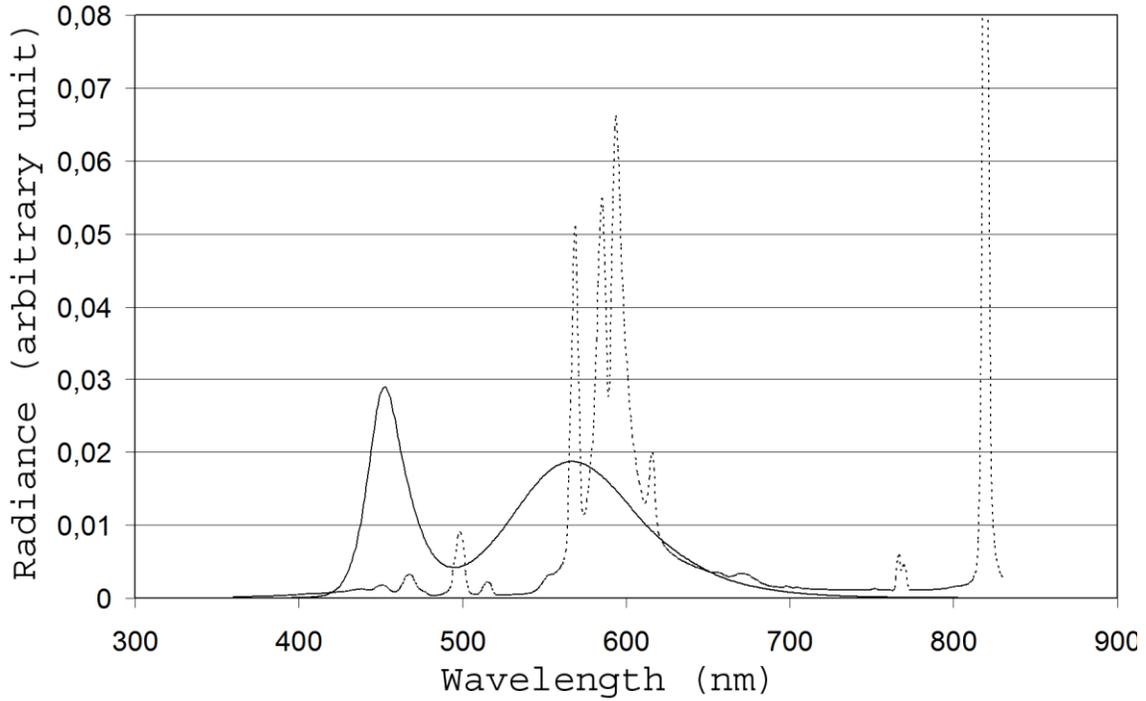

Figure 5. Spectral power distributions of a white LED (solid line) and a HPS lamp (dotted line) with equal photopic lumen output.

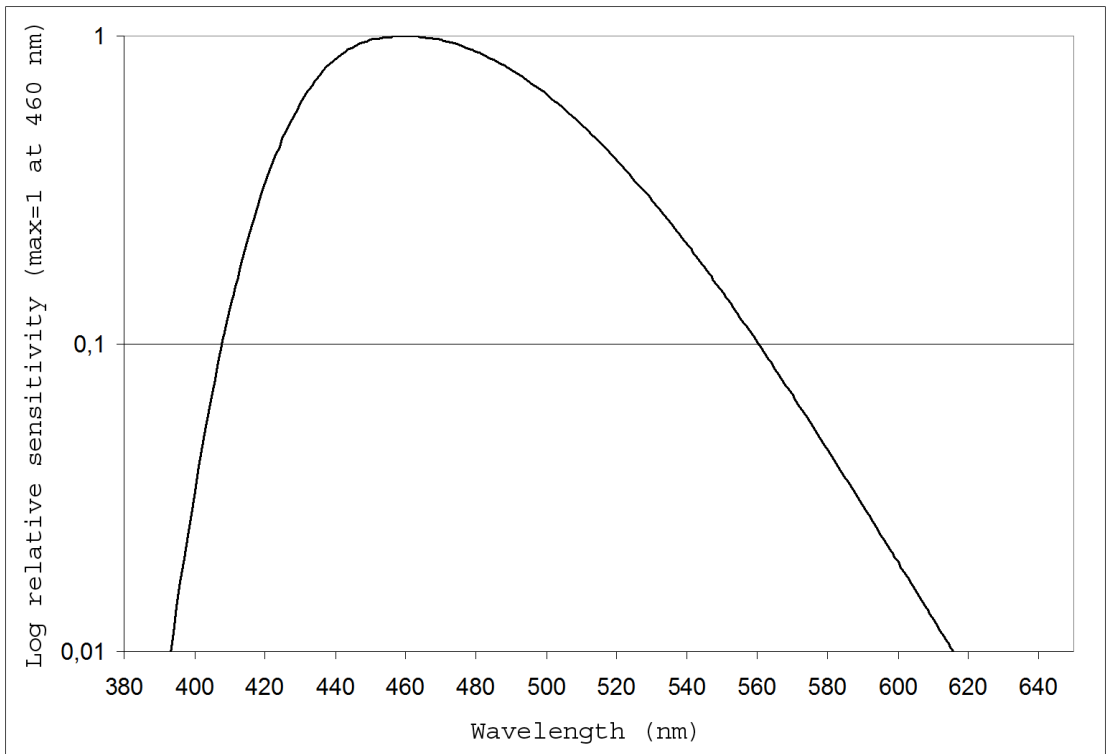

Figure 6. Action spectrum of melatonin suppression by light [57]